\documentclass[12pt]{iopart}
\usepackage{epsf}
\begin{document}

\title[Decoherence in the Heisenberg model]{Decoherence in the Heisenberg model}

%% __________________________________________________________________________|
%   Useful notations in text
% ---------------------------
\newcommand{\work}{{\fbox{\tiny work}}}
\newcommand{\refok}{}                    % {{$\bullet$}}      % Check others
\newcommand{\dataref}{{$\bullet$}}                            % Check REF
%% __________________________________________________________________________|
% Hyphenation problems/difficult names
% ------------------------------------
\hyphenation{Dannen-berg}
\hyphenation{Natur-wissen-schaften}
\newcommand{\Sch}{{Schr{\"o}dinger}}
% ___________________________________________________________________________|
%% Do not leave blanks in here
%% __________________________________________________________________________| 
%\preprint{quant-ph/0202169}

%
\author{Olavi Dannenberg}
\address{Helsinki Institute of
Physics, PL 64, FIN--00014 Helsingin yliopisto, Finland}
\ead{olavi.dannenberg@helsinki.fi}
%\date{November 22, 2002}
\date{\today}

% __________________________________________________________________________|
\begin{abstract}
% __________________________________ Do not leave a blank line here! _______|
We study a simplified Heisenberg spin model in order to clarify
the idea of decoherence in closed quantum systems. For this purpose,
we define a new concept: the decoherence function $\Xi(t)$,
which describes the dynamics of decoherence in the whole system, and
which is linked with the total (von Neumann) entropy of all
particles. As expected, decoherence is understood both as a
statistical process that is caused by the dynamics of the system, and
also as a matter of entropy.
Moreover, the concept of decoherence time is applicable in
closed systems and we have solved its behaviour in the Heisenberg
model with respect to particle number $N$, density $\rho$ and spatial
dimension $D$ in a $1/r$ -type of potential. We have also studied the
Poincar\'e recurrences occurring in these types of systems: in an
$N=1000$ particle system the recurrence time is close to the order of
the age of the universe. This encourages us to conclude that
decoherence is the solution for quantum-classical problems not only in
practice, but also in principle.
\end{abstract}
% ___________________________________________________________________________|

\submitto{\JPA}
%\keywords{decoherence, decoherence time, finite (closed) quantum system}
\maketitle

\section{Introduction}

Decoherence is widely accepted as an explanation of how quantum
correlations are damped out to make physical systems effectively
classical. Open (infinite) quantum
systems have been studied in great detail by many researchers
\cite{cl83,UZ89,zu91,hpz92,alzp95}. Their works 
are not relevant in our case, because they consider {\it an infinite
environment} that consumes the quantum coherence irrevocably. But how
about a finite environment or a {\it finite system} without an environment? In 
principle, if time is unlimited a finite system returns arbitrarily
close to its starting position an infinite amount of times (Poincar\'e
recurrence). Therefore, if a finite system starts from a superposition state
it begins to lose its coherence, but will at some moment return back
to its initial superposition state. Within these systems, is it
reasonable to talk about decoherence, and whether there is  some kind of {\it
preferred}, i.e., {\it pointer basis} as Zurek calls it \cite{zu82},
that is realised by decoherence. If the answer is
``yes'', is it somehow possible to circumvent decoherence in closed
quantum systems?

Our interest in closed and finite quantum systems arises from the
``cosmological'' aspects of reality. The universe has no
environment \cite{zu91} and it has a finite number of degrees of
freedom \cite{ll02}. Yet decoherence is observed in our
universe \cite{br96}. To understand decoherence, one should be able to
model these critical aspects of the
universe in decoherence studies. Previously closed quantum systems
have been studied \cite{dh92,gmh93,bh99} using the frame of the many histories 
interpretation of quantum mechanics. This approach is, however, found
to be problematic \cite{dk96,k96,k97}. In this study we consider the
off-diagonal elements of a reduced
density matrix to avoid the problems of many histories. 
Another reason is that decoherence theory using reduced density
matrices has not been studied in great detail, unlike the decoherent
histories approach. A well established
decoherence theory using reduced density matrices may clarify the
concept of decoherence in closed systems. Moreover, these approaches
are not equivalent \cite{gmh93}.

Elsewhere we have analysed in detail the conceptual
problem of decoherence in closed and finite systems \cite{da03}. A few
major results should also be outlined here, since, they are the key
concepts and premises.
\begin{enumerate}
\item There are two different decoherence types (similar to the
different entropy types): the {\it idealistic} and the {\it realistic}
decoherence. 
\begin{itemize}
\item The idealistic decoherence scheme can be applied only by those
observers who do not interact with the universe, and
who know the wave function of the universe and its time
evolution. In a closed system, there is no idealistic decoherence,
since the wave function of the universe remains always pure.
 
\item The realistic decoherence scheme is the internal view of the
universe
calculated from the wave function of the universe. It describes the
events as the real observers that are totally correlated with the
universe perceive them. To acquire this realistic viewpoint, one
should make an effective theory from the wave function of the
universe, e.g., to use reduced density matrices. This is often
referred to as coarse-graining.
\end{itemize}
In this study, the word ``decoherence'' refers to the
concept of realistic decoherence. Exceptions are mentioned.

\item The possibility of {\it recoherence} does not mean that there is
no decoherence. Decoherence is {\it the decay of the off-diagonal elements
of the reduced density matrix}, and hence, {\it recoherence means the
growth of the off-diagonal elements of reduced density matrix}. All
finite quantum systems may experience recoherence.
\end{enumerate}

Idealistic decoherence is often referred to with the word
``decoherence'', since, in open and infinite system studies both
decoherences behave similarly. The principal difference between
idealistic and realistic decoherence can be seen only in closed systems.

Using a reduced-density-matrix approach, we have studied a Heisenberg
spin model in order to clarify the idea of decoherence in closed
systems. We focus on one particle coherence (and entropy), and
thereby calculate the time evolution of the system. This
coarse-graining makes our (closed) system effectively open.   
Within this model, it is easy to
sketch how decoherence is advancing in the system. The main task of
this research is to derive functional dependences of decoherence time
on relevant parameters of the system (particle number $N$, density $\rho$,
potential and dimension $D$ of the system), i.e., to determine how the
decoherence time depends on these system variables, and why.

This paper consists of four main sections. First, we introduce our model
in \sref{model}. In \sref{theory} we present theoretically the
dynamics of a simple initial state, and define a decoherence function
$\Xi(t)$ as the measure of the coherence of the whole system.
\Sref{time} explains the
structure of our simulations, along with the main results. 
\Sref{discussion} is for discussion.

\section{The model \label{model}}

We have chosen the Heisenberg spin model because it is simple enough to
solve, and yet complicated enough to simulate properties
of real quantum systems. Coupled spin systems are interesting from a
quantum computational point of view, too. Our system, $N$ interacting
particles fixed in space, has no environment and, in that
sense, the system forms a closed quantum universe. The particles are spin-$1\over{2}$
particles, and the interaction between them is due to their spin
$z$-component (analogous to the Ising model). When there is no coupling
with the environment (i.e., no outer environment), the two spin states
have the same energy, which is taken to be zero. Zurek \cite{zu82} and
Omn\`es \cite{om94} have considered a similar, but simpler model in
order to show that off-diagonal elements
(i.e., quantum correlations) will decay in time. They labeled one
particle as {\it the system}, and the others as {\it an environment}
\footnote{In their model the particles that form an environment do not
interact with each other.}; but, we study the particle system as a whole.

The interaction Hamiltonian,
\begin{equation}
H=\hbar\sum^{N-1}_{j=1}\sum^{N}_{i=j+1}g_{ij}\sigma^{j}_{z}\otimes
\sigma^{i}_{z}\prod^{N}_{k=1, \ne j,i}\otimes\mathbf{1}_{k},
\label{f0}
\end{equation}
describes the dynamics of the system. The interaction matrix $G$, where
$g_{ij}=g_{ji}$, gives the interaction
strength between particles $i$ and $j$. The interaction strength
arises from the potential $V$; but, for formal calculations there is no
need to know more about it, because particles are doomed to stay in
one place. Fixing the positions of the particles 
is a justified assumption in decoherence
studies since, in most cases, the decoherence time scale is the shortest time
scale \cite{UZ89}, at least shorter than the time scale of particle motion.

\section{Theoretical calculations \label{theory}}

Let us now consider only the simplest case in order to present our
method, namely initially a product state of superposition states
\begin{equation}
\vert \Psi(0)\rangle=\prod^{N}_{k=1}\otimes\left(
a_{k}\vert+_{k}\rangle+b_{k}\vert-_{k}\rangle\right),
\label{f1}
\end{equation}
where $a$'s and $b$'s are normalised probability amplitudes $\vert
a_{k}\vert^{2}+\vert b_{k}\vert^{2}=1$ for all
$k=1,\dots,N$. The Schr\"odinger equation,
\begin{equation}
\rmi\hbar\partial_{t}\vert \Psi(t)\rangle=H\vert \Psi(t)\rangle,
\end{equation}
gives the dynamics of the system, and with the given initial condition
of equation (\ref{f1}) one gets the time dependence
\begin{equation}
\vert \Psi(t)\rangle={\rm exp}\left[-\rmi\sum^{N-1}_{j=1}\sum^{N}_{i=j+1}g_{ij}
\sigma^{j}_{z}\sigma^{i}_{z}t\right]\prod^{N}_{k=1}\otimes
\left(a_{k}\vert+_{k}\rangle+b_{k}\vert-_{k}\rangle\right).
\end{equation}

The fate of the $l^{{\rm th}}$ particle is solved
by tracing over other particles, i.e., degrees of freedom,
\begin{equation}
\rho_{l}={\rm Tr}_{1,\dots,N \ne l}\rho,
\end{equation}
where $\rho=\vert\Psi(t)\rangle\langle\Psi(t)\vert$. This is a
crucial step. We make an effective theory of our particle system by
tracing over the ``uninteresting'' particles (that form an effective
environment to the particular particle of our interest), as in the mean field
approximation. The net effect of traced-out particles is described in a
simpler form and with lesser degrees of freedom. This results in the
particular particle under consideration being an effectively open
system. The validity of this type of coarse-graining can be checked
by comparing the results with the entropy studies
(see \sref{recurrence}).

We thus have
\begin{eqnarray}
\rho_{l}&=\vert a_{l}\vert^{2}\vert+_{l}\rangle\langle+_{l}\vert+ \vert
b_{l}\vert^{2}\vert-_{l}\rangle\langle-_{l}\vert \nonumber \\
&+ \left[a_{l}b^{*}_{l}\prod^{N}_{k=1,
\ne l}\left(\vert a_{k}\vert^{2}\rme^{-\rmi 2g_{lk}t}+\vert
b_{k}\vert^{2}\rme^{\rmi
2g_{lk}t}\right)\vert+_{l}\rangle\langle-_{l}\vert + h.~c.\right].
\label{f2}
\end{eqnarray}
It is interesting that the result of equation (\ref{f2}) is the same as
in reference \cite{zu82}, if one drops the index $l$ away. In
reference \cite{zu82} only an interactionless environment has been
considered, but our model
counts all the interactions between particles. The result of equation
(\ref{f2}) is obvious: one particle of the system will be a victim
of decoherence. In fact, all particles separately will be victims of
decoherence. Other particles act as an environment for the particle of
interest, and the
coherence of this particular particle is dumped (i.e., displaced)
temporarily into its ``environment''. Now the interesting question
concerns the
decoherence of the whole particle system, as opposed to the
decoherence of the particles separately.

\begin{figure*}[ht] % _______________________________________________________%
\begin{center}
\epsfbox{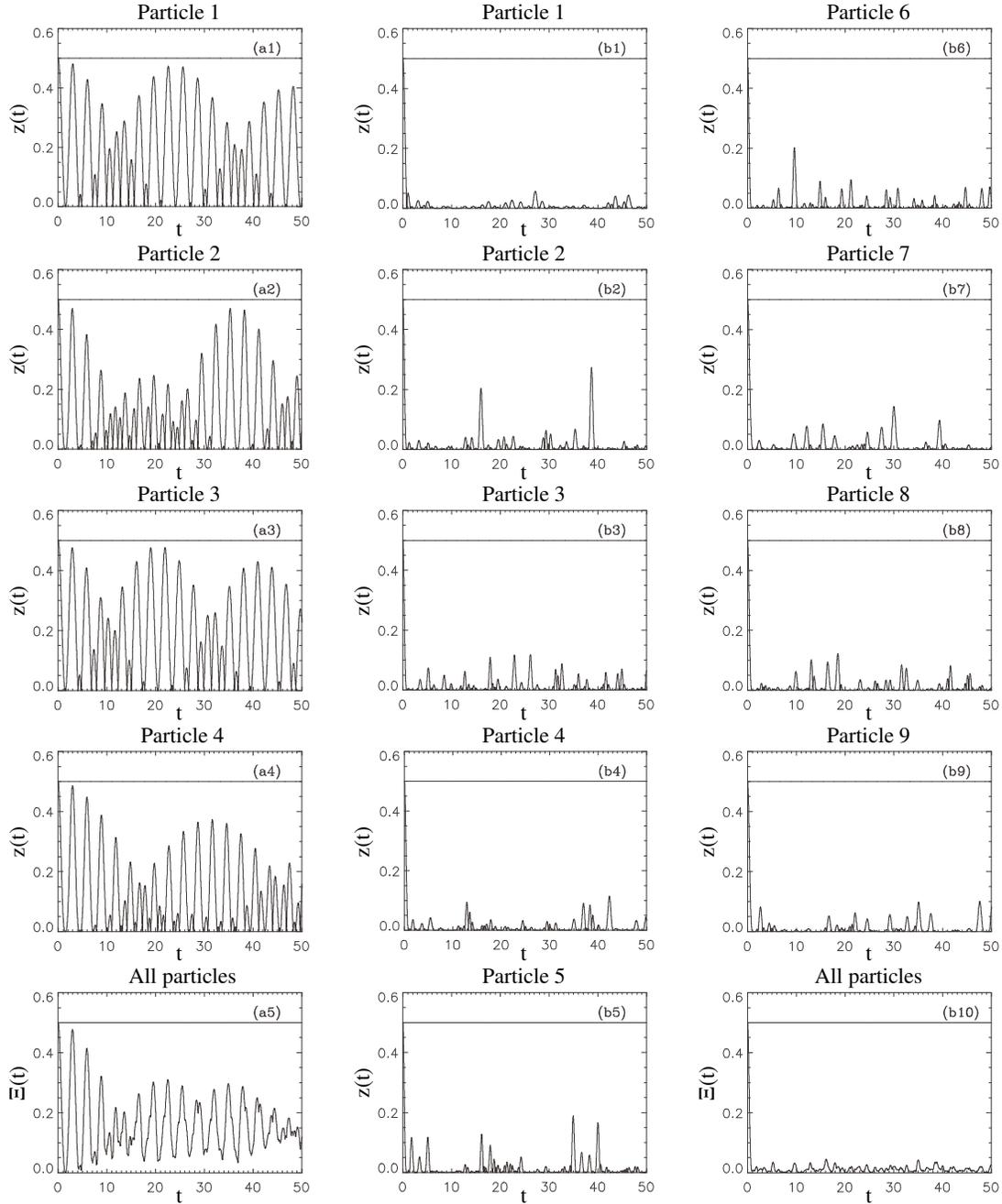}
\end{center} 
\caption[]{We illustrate two cases with different $N$: (a), 
$N=4$ and (b), $N=9$. For all particles $l=1,\dots,N$, $z_{l}(t)$ is
plotted. We have also plotted
$\Xi(t)={1\over{N}}{\sum^{N}_{l=1}\vert z_{l}(t)\vert}$ (a5, b10) which 
describes the decoherence of the whole system. Straight lines
represent the initial states. For $N=4$,
recurrences are stronger and occur more often than for $N=9$ (compare
(a5) and (b10)).}
\label{fig1}
\end{figure*}

The answer can be reasoned out in the following way: let us first make our
notation a bit lighter by denoting
\begin{equation}
z_{l}=a_{l}b^{*}_{l}\prod^{N}_{k=1,
\ne l}\left(\vert a_{k}\vert^{2}\rme^{-\rmi 2g_{lk}t}+\vert
b_{k}\vert^{2}\rme^{\rmi 2g_{lk}t}\right).
\end{equation}
This $z_{l}$ (or its complex conjugate $z^{*}_{l}$) describes the fate of
the off-diagonal elements of $l^{th}$ particle. Let us put the whole
system into the superposition state $a_{k}=b_{k}={1\over{\sqrt{2}}}$, 
$\forall k$, and let the elements of the matrix $G$ be random
numbers at the interval $[0,1]$ \footnote{This means that the interaction
between the particles does not depend on the distances between
particles, so, in this case, adding particles into the system means
the same as increasing the density of the system with $r$-dependent
interactions. It is obvious that decoherence is density-dependent, and
thus we will consider more realistic interactions in section \ref{time}.}.
The time dependencies of the off-diagonal elements for all particles
of the system are presented in \fref{fig1}. Note that parts of
the system may 
come close to their starting level, but each recurs at different times. It is
obvious that the whole system returns to its starting point more
seldom than its parts, so the fate of all off-diagonal elements of the
system is described by the function
\begin{equation}
\Xi(t)={1\over{N}}{\sum^{N}_{l=1}\vert z_{l}(t)\vert}.
\end{equation}
We have returned to a description of the whole system; but, $\Xi(t)$ is a
quantity of an effective theory, and therefore our initially closed
system becomes effectively open. We are definitely describing the
system from the inside view of the system, i.e., using the realistic
decoherence scheme.

If $\Xi(t)$ achieves its initial level, the system has
returned to its initial position, i.e., superpositions have
recurred. The complement event of $\Xi(t)$,
i.e. $Y={1\over{N}}\sum^{N}_{k=1}\vert a_{k}b^{*}_{k}\vert-\Xi(t)$,
behaves as the statistical entropy in classical physics: it grows fast
to some equilibrium value with certain fluctuations that depend on the
number of particles in the system [in figure \ref{fig4} we have
presented the behaviour of $\Xi(t)$]. In open and infinite
systems entropy and decoherence are related to each other (see, e.g., reference
\cite{UZ89}), and therefore it is reasonable to assume that the same
holds also in closed (and finite) systems. The decoherence function,
$\Xi(t)$, is a quantity similar to the sum
of all one particle (von Neumann) entropies of the system. 
The most stable (and the most probable)
state is the one with the maximum entropy, i.e., with minimum quantum
coherence. The system tends to reach this state and to spend most of its
time in it. It should be clear that decoherence is a similar
dynamical process as the growth of the entropy (see
\sref{recurrence}); therefore, recurrences in $\Xi(t)$ are rare
and short-termed phenomena.

Our model shows that decoherence is advancing at different speeds at
different parts of the system. The decoherence speed of the 
whole particle system differs from decoherence speed of the parts of
the system, and it can be evaluated from the normalised sum of all
particles, i.e., $\Xi(t)$. Also, the pointer basis is realised. In our
model with a particular interaction of equation (\ref{f0}), the pointer
basis is the $\sigma_{z}$-basis. It is quite reasonable to talk
about decoherence (and entropy) in a closed system.

\section{Decoherence time \label{time}}

In infinite quantum systems, decoherence time is easily defined: it is
the time when off-diagonal elements have decayed by the factor
$\rme^{-1}$ (e.g. \cite{UZ89,zu91}). If this definition is
straightforwardly applied in closed systems, the following problem will
appear. Decoherence is advancing at different speeds at different parts
of the system, so what is the decoherence time in this case? In open
systems the decay of off-diagonal elements nicely follows
the function $\rme^{-t/\tau_{d}}$, where $\tau_{d}$ is decoherence time; but,
in closed systems the decay function fluctuates, in some cases
fluctuating a lot
compared to the usual exponential behaviour. What is the decoherence time
when these fluctuations are present? 

We have solved this problem in the following way. First, we focus on
the
decoherence time of the {\it whole} system. Therefore, we use
$\Xi(t)$ to solve for 
the decoherence time. Of course, it is also possible to pay attention only to
some subsystem $S$, and solve the behaviour of this subsystem using
\begin{equation}
\Xi_{S}={1\over{m}}\sum^{m}_{l \in S}\vert z_{l}(t)\vert.
\end{equation}
Second, we use a least squares fit on $\Xi(t)$ using a function
$\xi(t)=(0.5-c)\rme^{-t/\tau_{d}}+c$ that
allows fluctuations around the average level $c$. This
fluctuation model is valid only when $t$ is small. ``Real'' time averages
and fluctuations are calculated after $\Xi(t)$ has stabilised. The
above mentioned decay time $\tau_{d}$ is the decoherence
time. By the way, the given definition yields the same results in open and
infinite systems as the familiar form ${1\over{2}}\rme^{-t/\tau_{d}}$,
because in open and infinite systems $c\rightarrow 0$.

\subsection{Numerics and simulations \label{numerics}}

Our main task is to try to derive the function
$\tau_{d}=\tau_{d}(N,\rho,V,D)$. The starting point in our
simulations is a $D$-dimensional ``box'' whose volume is
$l^{D}$. $N$ particles are placed randomly in this box. These
particles are in fixed places, and they interact with each
other according to the Hamiltonian (\ref{f0}). In this paper, we
concentrate on potentials of type
\begin{equation}
g_{ij}={\eta\over{\vert\bar{r}_{i}-\bar{r}_{j}\vert^{\epsilon}}},
\label{g0}
\end{equation}
where $\epsilon=1$. Now we calculate $\Xi(t)$ and fit it to the function
\begin{equation}
\xi(t)=(0.5-c)\rme^{-t/\tau}+c.
\end{equation}
We also solve the average level of $\Xi(t)$ by calculating its
time average over the interval $[t_{1},t_{2}]$, where $\tau_{d}\ll
t_{1}$ and $\tau_{d}\ll t_{2}-t_{1}$:
\begin{equation}
\langle \Xi\rangle={1\over{t_{2}-t_{1}}}\int^{t_{2}}_{t_{1}}\Xi(t)dt.
\end{equation}

This procedure is repeated $U$ times, after which we have a statistical
hunch about what is going on in the box. Finally, we average acquired values
of $\tau_{u}$ and $\langle \Xi\rangle_{u}$ to get a statistical
estimate. $\tau_{d}={1\over{U}}\sum_{u=1}^{U}\tau_{u}$ is a
function of parameters $N$, $\rho$, $V$ and $D$. Standard
deviations $\sigma_{\tau_{d}}$ and $\sigma_{\langle \Xi\rangle}$ are
also interesting. $\sigma_{\tau_{d}}$ is used in calculating error
estimates of our model, and $\sigma_{\langle \Xi\rangle}$ describes
fluctuations of $\Xi(t)$ around its average level $\langle
\Xi\rangle$. 

If the density under consideration is constant, then, when the
number of particles is changed, the size of the box is changed, too:
\begin{equation}
l=\left({N\over{\rho}}\right)^{1/D}.
\end{equation}

This simulation is repeated with different values of parameters
$(N_{i},\rho_{i},V_{i},D_{i})$, and we get results
$({\tau_{d}}_{i},\sigma_{{\tau_{d}}_{i}},\langle
\Xi\rangle_{i},\sigma_{\langle \Xi\rangle_{i}})$. The next task is to
find the  possible underlying functional dependence.

\subsection{Dependence on relevant parameters \label{dependence} }

We have focused on an initial state that contains complete
superpositions, so $a_{i}=b_{i}={1\over{\sqrt{2}}}$ \footnote{We want
to study a closed system with maximum initial coherence and minimum
initial entropy, and therefore, we set
$a_{k}=b_{k}={1\over{\sqrt{2}}}$. Of course, the choice of $a_{k}$
(and $b_{k}$) is arbitrary and we have tested that the results with
$a_{k}$'s as random complex numbers $[0, \vert a_{k}\vert^{2}=1]$
behave similarly as in the case $a_{k}=b_{k}={1\over{\sqrt{2}}}$.}. The
number of
simulations with particular parameter values is $L=100$. Entangled
particles have not (yet) been studied in relation to the case of
determination of decoherence time, so our initial state is the state
of equation (\ref{f1}). The time average is calculated in the interval of
$t_{1}=50$ to $t_{2}=100$. 

In this paper, we only consider the potential (\ref{g0}) with
$\epsilon=1$. The average level of $\Xi(t)$
is only a function of $N$, and it behaves as
\begin{equation}
\langle \Xi\rangle(N)=A\rme^{-BN},
\end{equation}
where $A\sim 1.27$ and $B\sim 0.43$. The fluctuation level around
$\langle \Xi\rangle$ is the same type of function: $\sigma_{\langle
\Xi\rangle}(N)=F\rme^{-GN}$, where $F\sim 0.03$ and $G\sim 0.3$.

Zurek \cite{zu82} and Omn\`es \cite{om94} have reported
that theoretical fluctuations of one particle coherence, $z(t)$, are $\sim
\rme^{-N\ln{2}/2}\simeq \rme^{-0.3466N}$, so it is reasonable that our
numerical results for $\Xi(t)$ behave in a similar manner.

The behaviour of decoherence time is more complicated, but it can also be
analysed. The function which fits well to simulated data is
\begin{equation}
\tau_{d}={1\over{\eta}}\left({P\over{N^{Q}}}+R\right)\rho^{-S},
\label{dekotime}
\end{equation}
where $P,Q,R,S$ are fitting constants which depend on dimension
$D$, and where the potential scaling factor $\eta$ is taken
into account. In Table \ref{tab1} we present the values of these
fitting constants in dimensions $D=1,2,3$. Dimensional dependencies
may be extracted from these results.

Some general results for different kinds of systems can be calculated.
For $D=3$ system of $N=100$ atoms in
$V=0.01~{\rm m^{3}}$, whose interaction strength is an
electromagnetic-type interaction ($\eta\simeq 2.2\cdot 10^{6}$), the
decoherence time is $\tau_{d}\sim 2\cdot 10^{-10}~{\rm s}$. 
We encourage experimentalists to do experiments
with as good as possible isolated closed interacting quantum
systems.

In \fref{fig4}(a2, b2, c2) differences between the dimensions are
considered. The analysis shows clearly that $\Xi(t)$ describes the
decoherence of the system quite well; especially when $D>1$, it follows nicely
the exponential form that is typical for open and infinite quantum systems.

\subsection{Recurrence \label{recurrence} }

The interesting thing to notice is that our quantum system (of $N$
particles) is {\it a closed and finite quantum system}. That means,
roughly speaking, that quantum correlations are never lost. They are
only displaced, and the system may return to its initial state, if
one waits long enough. But how much time is long enough? The
answer can be reasoned out in the following way. Let us give an approximation
about {\it Poincar\'e recurrence time} $T_{P}$ of our almost periodic
function $\Xi(t)$. In reference \cite{GA98} it has been argued
that the period of a function of type
\begin{equation}
F=\sum_{i} \cos{\omega_{i} t}
\end{equation}
is
\begin{equation}
T_{P_{F}}={2\pi\over{{\rm min}\vert\omega_{j}-\omega_{i}\vert}}.
\end{equation}
This is quite elementary. The same line of thinking is
applicable to the products of cosine functions, too. So, we can give an estimate of
period $T_{P_{\Xi}}$ for the function of type
\begin{equation}
\Xi(t)={1\over{N}}\sum_{i=1}^{N}\prod_{j=1, \ne
i}^{N}\cos{2g_{ij}t},
\end{equation}
that is,
\begin{equation}
T_{P_{\Xi}}={2\pi\over{{\rm
min}\vert\omega_{i}-\omega_{i'}\vert}},
\end{equation}
where $\omega_{i}=2{\rm min}\vert g_{ij}-g_{ij'}\vert$.

The recurrence time in our simulations can be solved, and in Table
\ref{tab2} some examples with various $N$ and $D$ are given. The
effect of $\rho$ shows up only as a common factor of $g_{ij}$ in
$\omega_{i}$. 

\begin{figure*}[ht] % ________________________________________________________%
\begin{center}
\epsfbox{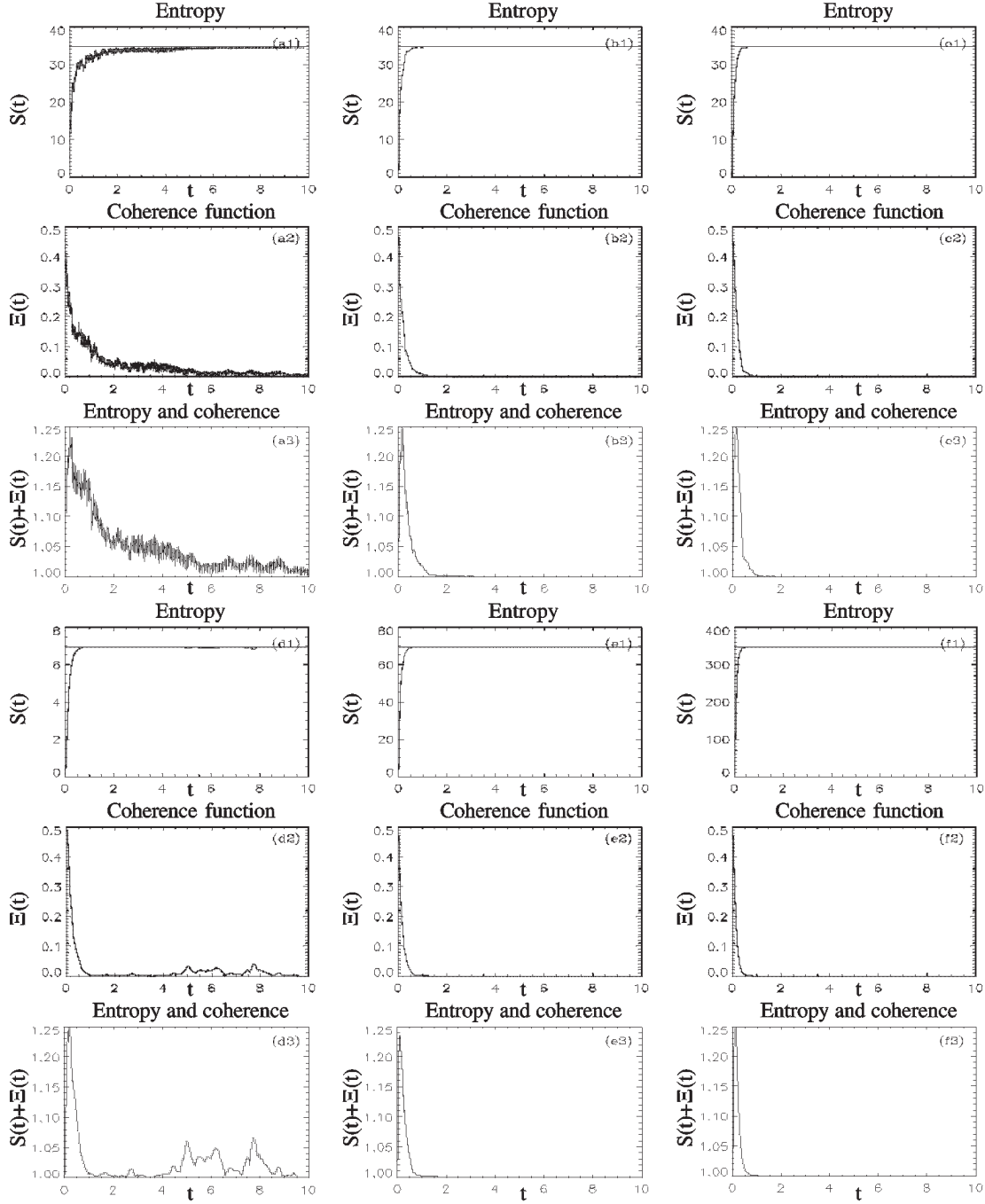}
\end{center} 
\caption[]{These figures illustrate the relation between entropy
$S(t)$ (1) and
coherence function $\Xi(t)$ (2). Various cases are
considered (with $\rho=1$): (a) $N=50, D=1$; (b) $N=50, D=2$; (c)
$N=50, D=3$; (d) $N=10, D=3$; (e) $N=100, D=3$; (f) $N=500, D=3$. It
seems clear that entropy (1) and coherence function (2) are
mirror images. Also the sum ${1\over{2N\ln{2}}}S(t)+\Xi(t)$ is plotted
(3).}
\label{fig4}
\end{figure*}

Let us study the possibility for recurrences from an entropy-based point
of view. The von Neumann entropy of a particular system that is
described by the density matrix $\rho$ is
\begin{equation}
S(\rho)=-{\rm Tr}(\rho\ln\rho).
\end{equation}
The entropy of a pure state is, of course, equal to zero. Information
that lies in correlations between particles is given by
\begin{equation}
I=\sum_{l=1}^{N}S(\rho_{l})-S(\rho),
\end{equation}
where index $l$ counts subsystems, e.g., particles (see reference \cite{pr98}).
In our system, $S(\rho(t))\equiv 0$ because the system is
closed and it starts from a pure state
$\vert\psi(t)\rangle\langle\psi(t)\vert$. But the entropy of subsystems
grows because the information, which is in the correlations between particles,
grows. Let us now put this into a quantitative form: the total entropy of
subsystems is 
\begin{equation}
S_{tot}=\sum_{l=1}^{N}S(\rho_{l})=-\sum_{i,l}\lambda_{i,l}\ln{\lambda_{i,l}},
\end{equation}
where $\lambda_{i,l}={1\over{2}}\pm{1\over{2}}\sqrt{1-4(\vert
a_{l}\vert^{2}\vert b_{l}\vert^{2}-\vert z_{l}(t)\vert^{2})}$ are the
eigenvalues of the reduced density matrix $\rho_{l}$. If $\vert
a_{l}\vert^{2}=\vert b_{l}\vert^{2}={1\over{2}}$, then
$\lambda_{i,l}={1\over{2}}\pm \vert z_{l}(t)\vert.$ In \fref{fig4}
we have plotted $S_{tot}$ with various particle numbers $N$. It is
notable that the total entropy behaves as a mirror image of
$\Xi(t)$. Moreover, the shape of the sum
$\Xi(t)+{1\over{2N\ln{2}}}S_{tot}(t)$ is similar in every case, even
if in $D=1$ there are large fluctuations present. The peak in figures
\ref{fig4}(a3-f3) indicates that, in the beginning, the entropy grows
slightly faster than decoherence, but, after the system is relaxed,
they are equal. Decoherence and entropy behave regularly; they are linked
to each other because they have the same origin (quantum dynamics). Our
analysis shows that $\Xi(t)$ is the right concept to describe
decoherence in closed and finite systems.

It seems clear that the system {\it may} return to its initial
position, but the recurrence time grows fast with respect to $N$. From
the entropy point of view, it is possible, in principle, for the system
to get completely to the initial state again, but, it is
thermodynamically impossible, i.e., the recurrence time is much greater
than the age of the universe.

\section{Discussion \label{discussion}}

We have studied this simple scenario in order to clarify the idea of
decoherence in closed systems. We have
stated that decoherence and entropy are two sides of the same
coin: they are well defined inside (effective theory) descriptions of
the closed system. This effective theory of a closed
system is valid for observers inside the universe, i.e., they observe
a truly closed and finite universe as effectively open. The universe
is truly closed from an outside point of view (complete description), but
the only way to access this view is an academic example. We leave the
detailed analysis of different types of observers in complete and
effective theories for elsewhere. Still, decoherence is a
working concept in closed systems also in principle, not only in practice,
e.g., as Bell has argued \cite{om94,be75}. Bell's
argument against decoherence has been refuted (e.g., by Omn\`es
\cite{om94}) \footnote{Bell's
argument is mainly that if the universe starts in a pure state, it
will always remain in a pure state, no matter how quickly the
off-diagonal elements of the reduced density matrix decrease and how
small they will become. He claims that this gives {\it in principle}
possibility to make such a measurement that will show quantum
interference. However, there is not even in principle such a
measurement device that can perform the measurement.

For a detailed discussion about the Bell's argument, one is encouraged
to study references \cite{da03,om94}.}. 

The possibility of recurrence makes neither concept,
entropy nor decoherence, empty. Decoherence is a dynamical process, arising
from interactions between particles, that diagonalises reduced density
matrices in the pointer basis. This holds true in our model.
Decoherence time $\tau_{d}$ can be applied and determined
in closed systems as well as in open systems, and the function $\Xi(t)$
describes the fate of all off-diagonal elements of particles forming
the system. $\Xi(t)$ can be linked with the entropy of the system as
a mirror image. The particles-in-a-box example in classical
statistical physics is analogous
to decoherence: the recurrence of all off-diagonal elements in our
system is a similar kind of phenomenon to the gas in a closed chamber
going completely to the other half of the chamber. This is possible,
but its possibility diminishes as $N$ grows. In the complete
description (view from outside), the total entropy and the idealistic quantum
coherence of a closed system are constants, but in the effective
theory (view from inside) sums of both realistic quantities are evolving in
time, as described in Sections \ref{dependence} and \ref{recurrence}.

It is quite obvious that the decoherence time $\tau_{d}$ behaves as $\sim
N^{-Q}$ and $\sim \rho^{-S}$, because decoherence is faster in bigger
systems (many particles and high density) than in small systems. An
interesting feature is that the decoherence time with respect to $N$
saturates to a certain value $R$ that is dimension-dependent. Using
equation (\ref{dekotime}) one can argue that an $N\sim 43000$ $3D$-particle
system behaves as an infinite particle system with an accuracy of
$99\%$. The amount of particles needed to cause a nearly-infinite
type of behaviour is small compared to the baryon number of our
universe ($\sim 10^{80}$ \cite{ll02}). The saturation effect gives
in principle a possibility to fight against decoherence: just make a
low density system. Of course, vacuum fluctuations are present in
low density cases, and therefore studies of the effects of second
quantisation are important in fighting decoherence. 

With our model, it is possible study the Schr\"odinger cat experiment
which crucially showed the problems in understanding
quantum mechanics and the demarcation problem between quantum and
classical \cite{sc35}. Many improvements have been made in quantum
mechanics since the days of Schr\"odinger, especially concerning
decoherence. This study covers basically the question of
quantum-classical problem; we have shown that quantum mechanics can be
applied to the whole universe, and as an effective theory we acquire
almost a classical universe with the help of decoherence. The
Schr\"odinger cat problem is more than this demarcation problem, but
the presented model can
be used in constructing a realistic enough model of a cat and its
surroundings. Entangled states will play a crucial role in that case.

The ``cat-in-the-box'' will be in our future interests, as well as other
$1/r^{\epsilon}$ potentials. Also, ideas how to circumvent
decoherence are interesting. The system geometry may play some
role in slowing down decoherence and making recurrence more
probable. Correlations between particles should be studied as well in
greater detail; here we have only considered an effective theory with
one-particle correlations, but many-particle correlations may
contribute to the dynamics of the system. However, our entropy studies
seem to justify the results. There is one minor drawback in
our model that is a result of simplifying and making an academic
example, and it may affect the results: the model is, strictly speaking,
discrete because the particles are placed (randomly, but this does not
matter) in fixed and 
{\it accurate} places. Clearly, this violates the Heisenberg uncertainty
principle and may cause considerable effects. It would be interesting
to study what kind of results are acquired with particles as
probability distributions in space.

%% __________________________________________________________________________|
\ack
The author would like to thank Kalle-Antti Suominen for guiding him into
the scientific world, and Magnus Ehrnrooth Foundation \& the Academy
of Finland (project 50341) for funding. He
also thanks John Calsamiglia for questions that have inspired him to
try to find answers, if there are any, and Matt Mackie for
constructive criticism.
%%b____________________________ no blanks here ______________________________|

%% __________________________________________________________________________|
% More than five authors, three first + et al. !!!!!!!!!!!!!!
% ___________________________________________________________________________|

\clearpage

\Tables
 
\begin{table}
\caption{\label{tab1}The values of fitting constants. Some constants
follow (within the error bars which are acquired from our statistical
analysis) a dimensional scaling which is presented on
the last row.}
\begin{indented}
\item[]\begin{tabular}{ccccc}
$D$ & $P$ & $Q$ & $R$ & $S$                            \\
\hline
1 & $3.73\pm 1.0$ & $1.49\pm 0.3$ & $0.415\pm 0.022$ & $1.80\pm 0.05$
\\
2 & $0.77\pm 0.3$ & $0.80\pm 0.1$ & $0.166\pm 0.004$ & $1.00\pm 0.05$
\\
3 & $0.45\pm 0.1$ & $0.55\pm 0.05$ & $0.128\pm 0.003$ & $0.67\pm 0.05$
\\
\hline
Scaling & $3.5/D^{2}$ & $1.6/D$ & & $1.80/D^{0.88}$
\end{tabular}
\end{indented} 
\end{table}

\begin{table} 
\caption{\label{tab2}Examples of estimated recurrence times 
$T_{P_{\Xi}}$ with various initial conditions. We have also
presented standard deviations of these numerically estimated 
recurrence times. Note that $T_{P_{\Xi}}$ is highly dependent on 
$N$ and $D$. The unit of time is $gt$.} 
\begin{indented} 
\item[]\begin{tabular}{ccccc}
$N$ & $\rho$ & $D$ & $T_{P_{\Xi}}$ & $\sigma_{T_{P_{\Xi}}}$   \\
\hline 
10   & 1  & 1 & $5.0\cdot 10^{6}$  & $1.1\cdot 10^{8}$         \\
50   & 1  & 1 & $4.6\cdot 10^{10}$ & $2.7\cdot 10^{11}$        \\
100  & 1  & 1 & $2.2\cdot 10^{12}$ & $1.2\cdot 10^{13}$        \\
500  & 1  & 1 & $6.3\cdot 10^{16}$ & $3.6\cdot 10^{17}$        \\
1000 & 1  & 1 & $5.5\cdot 10^{17}$ & $1.9\cdot 10^{18}$        \\
1000 & 1  & 2 & $2.2\cdot 10^{15}$ & $4.7\cdot 10^{15}$        \\
1000 & 1  & 3 & $4.0\cdot 10^{14}$ & $8.6\cdot 10^{14}$        \\
1000 & 10 & 3 & $2.0\cdot 10^{14}$ & $4.5\cdot 10^{14}$        \\
\end{tabular}
\end{indented}
\end{table}

\clearpage

\section*{Figures and figure captions}

 % ______________________________________________________________%

 % ______________________________________________________________%

%% __________________________________________________________________________|
%% __________________________________________________________________________|
\end{document}